\documentclass{article}
\usepackage{graphicx}

\usepackage[usenames]{color}
\usepackage{makeidx}

\begin{document}

\title{\bf Entanglement in parametric down-conversion with phase mismatch at finite temperature }

\author{Ana M. Martins \\
Departamento de F\'{i}sica, Instituto Superior T\'{e}cnico, 1049-001
Lisboa, Portugal }

\date{\today}

\maketitle

\begin{abstract}
In this work the degree of entanglement of two-photon states generated by parametric-down conversion with a phase mismatch is computed in terms of the temperature of the nonlinear medium where the interaction takes place. The minimum interaction time, named {\it birth time of entanglement}, needed before entanglement is attained, is computed in terms of the temperature and of the phase mismatch. Our results show that, for each value of the phase mismatch, there is a critical temperature above which entanglement disappears. We are able to identify the regions of parameters, temperature and mismatch, where the states are entangled and where they are separable. We also propose a entanglement witness based in a dynamical invariant of the nonlinear interaction.

\end{abstract}

PACS number(s)  03.67.a, 42.50-p., 42.65.Lm


\newpage

\section{Introduction}

 A significant number of Quantum Computation and Quantum Information protocols use the quadrature amplitudes of the quantized electromagnetic field \cite{Braunstein1998}, \cite{Plenio2003, Braunstein2005}. The essence of all these protocols rely in the possibility of producing entangled photons in a nonlinear medium \cite{Ou1992}.
 
 The Parametric Down Conversion (PDC) \cite{LOUISELL1961, Rubin1994} is one of the most often used processes for production of entangled pairs of photons. However, as for many nonlinear optical processes, PDC is ruled by phase matching whose importance was early recognized \cite{Kleinman1968}, Theoretical treatment of a phase mismatch that occurs during the propagation in the nonlinear medium is complicated and several approximations have been made \cite{Bennink2010,Fedorov2008,Yao2012}, in order to relate the phase mismatch,  with the amount of quantum correlations in the spontaneous emission of photons.
 
At room temperatures the average number of photons at optical frequencies is very small (on the order of $10^{-40}$). However, in the microwave part of the spectrum the number of thermal photons initially present in the nonlinear medium can not be ignored and we expect that they will play a determinant role in the amount of entanglement that can be extracted from the interaction. To our knowledge a comprehensive study of PDC processes with a phase mismatch and at finite temperature of the nonlinear medium has not been addressed.     

In this paper we describe the PDC process with a phase mismatch with a simple model that is the quantum transcription of the classical PDC approach derived by Bloembergen \cite{Bloenbergen1998}. This model is valid when the interaction is weak, and when the complex amplitudes of the interacting waves vary slowly with the propagating distance in the nonlinear medium. Our purpose is to relate the joint effect of the temperature of the nonlinear medium and of the phase mismatch, with the degree of entanglement of the down-converted photon states. 

 To quantify the degree of entanglement in terms of the phase mismatch and of the temperature we use the logarithmic negativity \cite{Horodecki1998}. We show that for finite temperatures there is a minimum time of interaction needed, the {\it Birth Time of Entanglement} (BTE), before the system starts to be  entangled and that it varies with the temperature and with the phase mismatch. For a given phase mismatch, we compute the degree of entanglement with the temperature in order to obtain the critical temperature, above which the quantum state becomes separable. This study allows us to identify what are the regions of joint values of temperature and phase mismatch, where the states of the system are entangled and where they are separable. Finally, we propose a feasible entanglement witness that is directly connected with a dynamical invariant of the system.

 The plan of the paper is the following: In Sec.II we introduce the Hamiltonean of the system with non zero phase matching, and obtain the time dependent  bosonic operators. In Sec. III we compute the {\it birth time of entanglement} and the degree of entanglement in terms of the phase mismatch and of the temperature and, an entanglement witness is proposed. Finally in Sec.IV we present the conclusions.

\section{System dynamics }

The parametric down conversion is a process where an intense pump wave of frequency $\omega_0 $, propagates in a nonlinear medium and gives rise to two other modes of smaller frequencies  $\omega_1$ (the signal) and $\omega_2$ (the idler) obeying the energy conservation rule $ \omega_0 = \omega_1 + \omega_2$. When the interaction is weak, and when the complex amplitudes of the interacting waves vary slowly with the propagating distance so that they can be assumed approximately constant within a distance of a wavelength, it is possible to use the slowly varying envelope approximation derived by Bloembergen \cite{Bloenbergen1998} in a classical context, to describe the nonlinear interaction with a phase mismatch $\delta$. The quantum version associated to the classical Hamiltonean describing this process, is 
\begin{equation}\label{hamiltonean}
{\hat H} = \hbar \omega_1 \hat{a}_1^\dag \hat{a}_1 +     \hbar \omega_2\hat{a}_2^\dag \hat{a}_2 -    \hbar  g (e^{- i [(\omega_0   + \delta) t + \varphi  ]} \hat{a}_1^\dag \hat{a}_2^\dag+h.c.)
\end{equation}
where the intense pump mode is treated classically as a coherent undepleted field of complex amplitude  $\alpha_0 = | \alpha_0| e^{- i [ ( \omega_0 + \delta )t  + \varphi ] } $ and an arbitrary phase $\varphi $. Modes $1$ and $2$ are described by the bosonic operators $\hat{a}_1$ and $\hat{a}_2$. The interaction time $t$ is the propagation time in the nonlinear medium and the coupling constant $g $ is proportional to the second order susceptibility of the medium and to the modulus $ |\alpha_0|  $ of the pump wave.

Solving the Heisenberg equations of motion of the system (see the Appendix), we obtain the time dependent bosonic operators
 \begin{equation}\label{time1}
\hat{a}_1(\tau)= e^{-i \tau \omega_1^{\prime}(y) }  \left[ A ( x \tau) \,\  \hat{a}_{10} + B ( x \tau) \,\   \hat{a}_{20}^{\dag} \right]
\end{equation}
\begin{equation}\label{time2}
\hat{a}_2^{\dag}(\tau)= e^{i \tau \omega_2^{\prime} ( y)} \left[ B^{*} ( x \tau) \,\   \hat{a}_{10} +  A^{*}( x \tau) \,\   \hat{a}_{20}^{\dag} \right]
\end{equation}
where $\tau = gt$ is an effective dimensionless interaction time, $y=\delta / g$ is the dimensionless mismatch parameter with values in the interval $0 \leq  y < 1$, $\omega_j^{\prime} (y)=  \omega_j / g + y$ is the dimensionless effective frequency of oscillation of mode $j$ and $x=  (1-y^2)^{1/2}$. The subscript zero refers to initial time. The coefficients $A(x \tau)$ and $B(x \tau)$ are defined by

 \begin{equation}\label{coefficientA}
A( x \tau )=  \cosh ( x \tau )+  \frac{i y  }{x} \sinh ( x \tau) 
\end{equation}
\begin{equation}\label{coefficientB}
B(x \tau)= \frac{i}{x} \sinh ( x \tau)  
\end{equation}

Equations (\ref{time1}) and (\ref{time2}) extend the well known result derived in \cite{LOUISELL1961} to take into account the influence of the phase mismatch, and reduces to it when $y =0 $. Because the pump amplitude $ | \alpha_0|$, is treated as constant, the solution to the parametric interaction ceases to be valid when the  number of generated pair of photons is such that determines an appreciable depletion of the pump mode.

The argument of the hyperbolic functions is the squeezing parameter,  
\begin{equation}\label{squeezing}
r(x \tau) = \tau  x
\end{equation}
It depends not only on the dimensionless interaction time $\tau$, but also on the  mismatch. Making $\delta =0$ in the last equation we obtain the usual squeezing parameter $r= gt $ \cite{LOUISELL1961}. For a given interaction time the squeezing parameter decreases with increasing mismatch and we expect that the efficiency of the correlated pair production also decreases with $y$.

\section{Degree of entanglement}

In this work, we are interested in the amount of entanglement we can extract from the correlated pair production in terms of the phase mismatch and at finite temperature $T$.

In the spontaneous PDC emission the signal {\bf 1} and the idler {\bf 2} modes are initially in the vacuum state $ |0 \rangle_{j0}$ ($j=1,2$). The initial state of the composite system is the pure and separable state given by the tensor product $ | \psi \rangle_0 =  |0 \rangle_{10} |0 \rangle_{20}$.

When the nonlinear medium is in thermal equilibrium at temperature T, the initial state of the two modes is a separable mixed state given by the density operator ${\hat \rho}_0 = {\hat \rho}_{10}  \otimes {\hat \rho}_{20}$,
where $ {\hat{ \rho}}_{j0}  = \sum_{n_{j} =0}^{\infty} e^{- \beta_j n_{j} } | n_{j}\rangle \langle n_{j} |$
is the thermal field density operator of mode $j$, with $ \beta_{j} =\hbar \omega_{j} / (k_B T) $ and $k_B$ is the Boltzmann constant. 

The initial mean number of photons in mode $j$ is ${\bar n}_{j0} (T)=1/(e^{\beta_j -1})$. At room temperature and in the optical part of the electromagnetic spectrum ${\bar n}_{j0}<<1$, however, in the microwave part of the spectrum ${\bar n}_{j0}>>1$ and we cannot ignore the presence of thermal photons in the nonlinear medium. The initial average number of photons in the limit $T \rightarrow  0$ is ${\bar n}_{j0} =0$, and we recover the vacuum state.

 Both vacuum and thermal states are Gaussian states and since the Hamiltonian ${\hat H}$ is bilinear in the field modes, the overall output state is also Gaussian and the separability criterion \cite{Simon2000,Werner01} is completely characterized by the first and the second statistical moments of the quadrature phase operators $ \hat { q}_j  = \sqrt{\frac{\omega_j}{2 \hbar} }({\hat { a}_j } + {\hat {a}_j^{\dag}  } )$ and $\hat {p}_j =   \frac{1}{{\sqrt  {2 \hbar \omega_j}}}({\hat { a}_j  } - {\hat {a}_j^{\dag}  } )$, $ (j=1,2)$. These operators can be grouped  together in a vector of operators $  {\hat X}  \equiv  ( {\hat q}_{\bf 1} , {\hat p}_{\bf 1} , {\hat  q}_{\bf 2},  {\hat p}_{\bf 2}  )^T$.

The first and the second statistical moments of the quadrature phase operators, will be denoted, respectively, by the vector $ {\bar X } \equiv  ( \langle {\hat { x}_1 } \rangle , \langle {\hat x}_2 \rangle , \langle {\hat { p}_1 } \rangle,\langle {\hat { p}_2 } \rangle )$ and by the covariance matrix (CM) ${\bf \sigma } $, of elements $\sigma_{ij} \equiv  \frac{1}{2} \langle  {\hat { x}_i  } {\hat { x}_j } + {\hat { x}_j  } {\hat { x}_i } \rangle - \langle  {\hat { x}_i  } \rangle  \langle {\hat { x}_j } \rangle $. The first moments can be adjusted to zero by local unitary operations which leave invariant entropy and entanglement.
The time dependent covariance matrix $\mathbf {\sigma} (\tau)$ can be expressed in terms of the three $(2 \times 2)$ block matrices $ \mathbf { \alpha }(\tau)$, $ \mathbf { \beta }(t)$ and $ \mathbf { \gamma }(t)$
\begin{equation}\label{macrovar}
{\bf \sigma }(\tau) = \left( 
\begin{array}{cc}
\alpha (\tau)  & \gamma (\tau)  \\ 
\gamma^T(\tau)    & \beta (\tau) \\ 
\end{array}
\right)
\end{equation}
The diagonal blocks  $ \mathbf { \alpha }(\tau)$, $ \mathbf { \beta }(\tau)$ are the local CM of modes {\bf 1} and {\bf 2}, respectively. The off-diagonal block, $ \mathbf { \gamma }(\tau)$, encode the intermodal correlations (quantum and classical) between the two modes.

Hamiltonians that are quadratic in the bosonic operators, possess universal quantum invariants, i.e., certain combinations of variances which are conserved in time independently of the concrete form of coefficients of the Hamiltonian \cite{Serafini04}. These invariants exist due to the symplectic structure of the transformation relating initial and time-dependent values of the quadrature components operators.  Concerning the statistical moments, these universal invariants are \cite{Dodonov2005}
\begin{equation}\label{invariants}
{\cal I}_1 = \det {\bf \sigma} (\tau)\,\ ;  \,\,\,\,\,\,\      {\cal I}_2 = \det \alpha (\tau)+ \det \beta (t)+ 2 \det \gamma (\tau)
\end{equation} 
Defining the quantities 
\begin{equation}\label{st}
{\cal S} (\tau) = {\cal S}_0 + \frac{1}{2} (\det \gamma (\tau) - |\det \gamma (\tau)| )\,\ ;  \,\,\,\,\,\,\      {\cal S}_0= {\cal I}_1  -   \frac{1}{4}{\cal I}_2  +\frac{1}{16}
\end{equation} 
the entanglement criterion can be written in a simple form
\begin{equation}\label{criterion}
{\cal S} (\tau) < 0
\end{equation} 
For both vacuum and thermal initial states, we prove in the Appendix, that $
{\cal S} (\tau) = S_0 -  | \det \gamma (t)| $, with ${\cal S} _0= { \bar n}_{10}{ \bar n}_{20} ( { \bar n}_{10}+1) ( { \bar n}_{20}+1)$. For the initial vacuum state, ${ \bar n}_{10}= {\bar n}_{20}=0 $ and ${\cal S}_0=0$.

\subsection{ Birth time of entanglement}

Using the entanglement criterion (\ref{criterion}), we conclude that the two modes are entangled for interaction times $t_i$ obeying the inequality
\begin{equation}\label{ent}
|\det \gamma (\tau_i)| > {\cal S} _0
\end{equation}
where $|\det \gamma (\tau_i)| $ is an increasing function of  $\tau$. The instant $\tau_{\cal E}$, such that $|\det \gamma (\tau_{\cal E})| 
= {\cal S}_0 $ is the minimum interaction time needed before entanglement appears, for obvious reasons it will be named the {\it birth time of entanglement} (BTE).

For the vacuum initial state, $\tau> 0 \Rightarrow {\cal S} (\tau) >  0$, therefore the two modes are entangled since the very beginning of the interaction, this is, $\tau_{\cal E} =0$ for any value of phase mismatch $y$. On the contrary, for an initial thermal state, the condition (\ref{ent}) is attained only after a finite time of interaction, $\tau_{\cal E}$, has passed.

The {\it birth time of entanglement} depends on the temperature $T$, through the initial average number of photons ${ \bar n}_{10}(T)$ and ${ \bar n}_{20}(T) $. To compute the  initial average number of photons we assume two typical frequencies in the microwave region of the eletromagnetic spectrum: $\nu_1 =3.12 \times 10^{10} $Hz and $\nu_2 =2 \nu_1  $  and for the coupling constant we assume $g =\pi  \times 10^{-2} \nu_1$.

The dependence of the dimensionless BTE, $\tau_{\cal E}$, over the temperature is displayed in Fig.1, for the dimensionless frequencies $ { \bar \omega}_1= 2 \pi \nu_1 / g=200 ,\,\    { \bar \omega}_2=2 \pi \nu_2 /g= 400$ and for different values of the dimensionless mismatch parameter $y$: $y =0$ (triangles), $y =0.5$ (circles) and $y =0.9 $ (stars). We observe that, for a given mismatch parameter $y$, the BTE increases with the temperature although this growth slows down at higher temperatures. As we will show, only when the average number of correlated photon pairs is higher than a threshold defined by the uncorrelated initial thermal photons is the {\it break-even} for entanglement attained.

\begin{figure}[htb]
\centering
\includegraphics{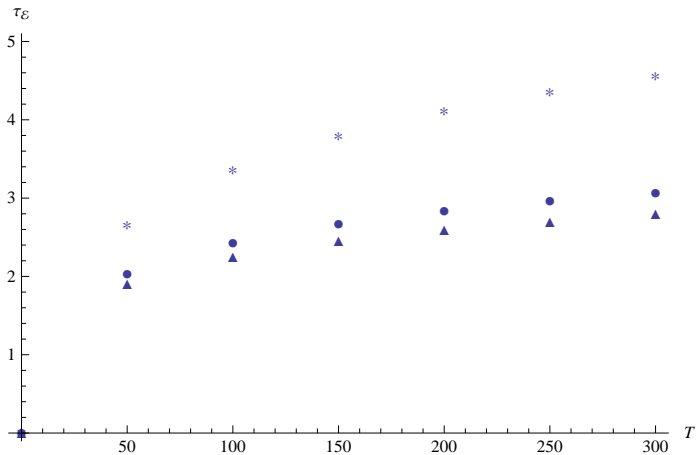}
\caption{\it 
Dimensionless BTE, $\tau_{\cal E}$, over the temperature, for the dimensionless frequencies $ { \bar \omega}_1 = \omega_1/g=200 ,\,\ {\bar \omega}_2=\omega_2/g= 400$ and for different values of the dimensionless mismatch parameter $y$: $y =0$ (triangles), $y =0.5$ (circles) and $y =0.9 $ (stars).}
\label{FIG.1}
\end{figure}

In Fig.2, the BTE is displayed as a function of the dimensionless mismatch parameter $y$, for different temperatures: $T=50^{\circ}K$ (triangles), $T=150^{\circ}K$ (crosses) and $T=300^{\circ}K$ (circles). We observe that, for $T > 0$, the BTE increases with $y$, and the growth rate is higher at higher temperatures. This is not surprising since the squeezing parameter, responsible for the efficiency of the photon pair production, decreases with increasing mismatch introducing a randomness in the interaction.

\begin{figure}[htb]
\centering
\includegraphics{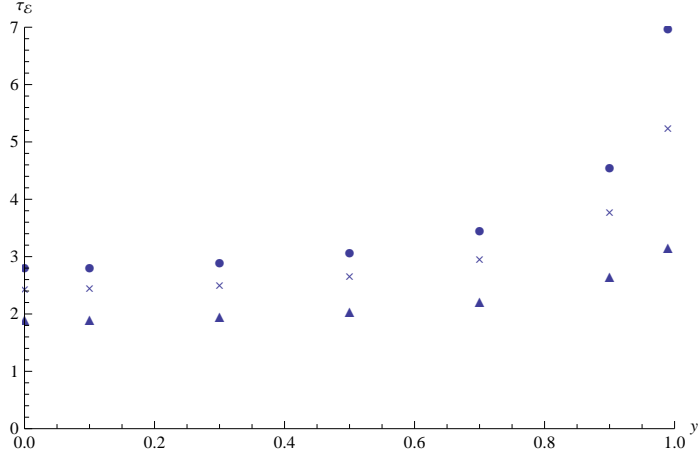}
\caption{\it 
BTE is displayed as a function of the dimensionless mismatch parameter $y$, for different temperatures: $T=50^{\circ}K$ (triangles), $T=150^{\circ}K$ (crosses) and $T=300^{\circ}K$ (circles). }
\label{FIG.2}
\end{figure}

\subsection{Logarithmic negativity}

The logarithmic negativity ${ \cal E_N}$ \cite{Horodecki1998, Werner2002} provides a proper quantification of entanglement in two-mode Gaussian states and is a computable measure of entanglement. 
 
In Fig.3 the degree of entanglement ${ \cal E_N}$ is displayed in terms of $y$ for $\tau_i = 4.543$ and for $T=0^{\circ}K $ (crosses), $T=50^{\circ}K$ (triangles) and $T=300^{\circ}K$ (circles). The same dimensionless frequencies as in Fig.2 are assumed.  We observe that, for a given temperature, the degree of entanglement decreases with increasing $y$. Moreover, for $T=300^{\circ}K$ and for $y \geq 0.9$, ${\cal E}_{\cal N}=0$, this is, the two modes are in a separable state though for $T=50^{\circ}K$ the system is still entangled for $y=0.99$. 

\begin{figure}[htb]
\centering
\includegraphics{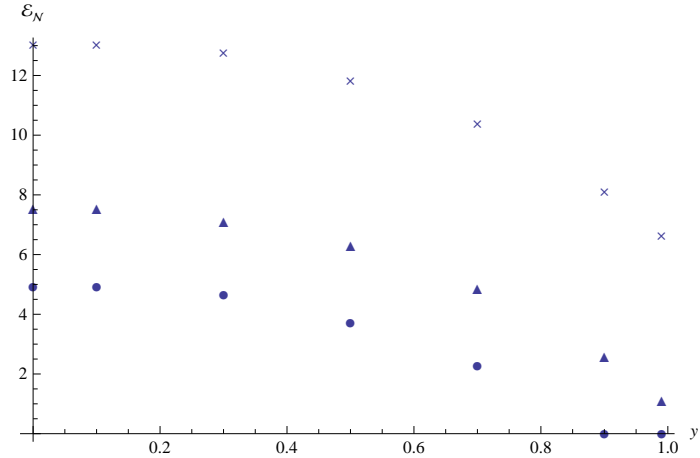}
\caption{\it 
Degree of entanglement ${ \cal E_N}$, in terms of $y$ for $\tau_i = 4.543$ and for $T=0^{\circ}K $ (crosses), $T=50^{\circ}K$ (triangles) and $T=300^{\circ}K$ (circles). The same dimensionless frequencies as in Fig.2 are assumed.}
\label{FIG.3}
\end{figure}

As a matter of fact, when the initial temperature of the nonlinear medium rises, there are initially more uncorrelated thermal photons in the modes. We expect intuitively that, for a given interaction time, any entanglement should vanish at some critical temperature $T_c$. We have computed ${ \cal E_N}$, for $\tau_i= 2.881$, in terms of the temperature and for different values of the mismatch parameter $y$, in order to find the critical temperature for each $y$. The results are displayed in Fig.4 where the points $(y, T_c)$ are marked with circles. The solid curve was obtained by interpolation. The quantum states of the two modes characterized by the parameters $y$ and $T$ in the shaded region under the curve are entangled states. On the contrary, above the curve these parameters define separable states. We observe that there is a kind of trade-of between the phase mismatch and the critical temperature: greater is the phase mismatch smaller is the critical temperature and vice versa. 

\begin{figure}[htb]
\centering
\includegraphics{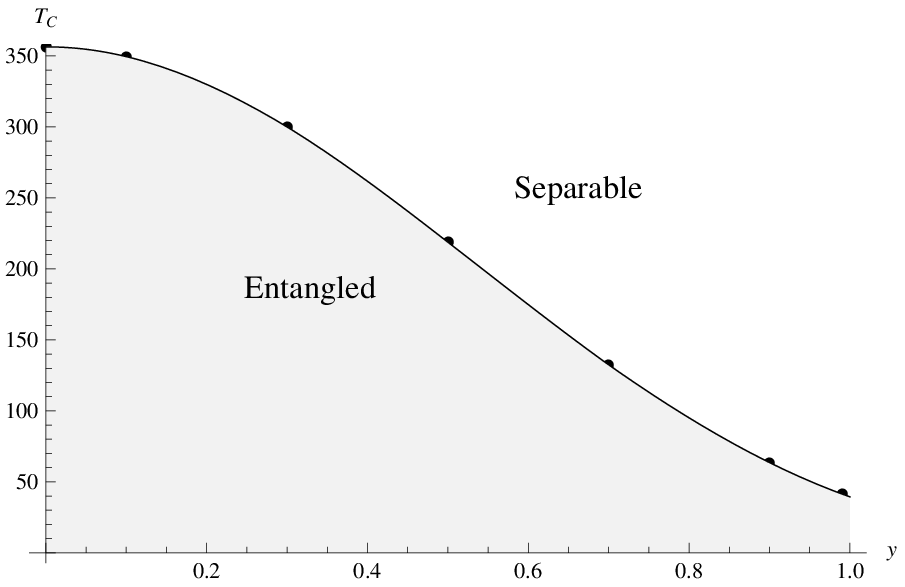}
\caption{\it 
 Critical temperature $T_c$ for different values of $y$ (circles). The solid curve was obtained by interpolation. The shaded region under the curve correspond to values of parameters $(T,y)$ where the states are entangled. Above the curve these parameters define separable states.}
\label{FIG.4}
\end{figure}

\begin{figure}[htb]
\centering
\includegraphics{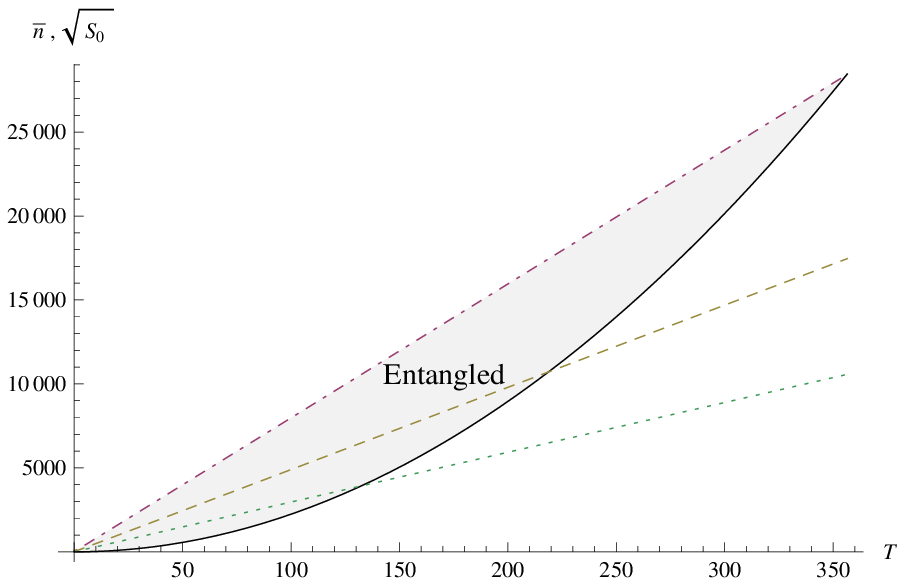}
\caption{\it 
Average number of photons $\langle {\hat n}(T)   \rangle = (\langle {\hat n}_1 (T)  \rangle  + \langle {\hat n}_2 (T) \rangle )/2$ produced in the PCD process after a time of interaction $\tau_i= 2.881$, in terms of the temperature $T$ for different values of the mismatch parameter: $y=0$ (dashed-dot line), $y=0.5$ (dashed line) and $y=0.7$ (dotted line). The invariant $\sqrt{S_0} $ (solid line) is displayed in terms of the temperature $T$. }
\label{FIG.5}
\end{figure}

Another interesting conclusion is that it is possible to have entangled states at room temperature for small enough phase mismatch. For instance, in the case considered in the Fig.4, the modes are entangled at $T=300^{\circ}K$  for $\delta \leq 0.3$.

\subsection{Entanglement witness}

In Fig.5 is displayed the average number of photons $\langle {\hat n}(T)   \rangle = (\langle {\hat n}_1 (T)  \rangle  + \langle {\hat n}_2 (T) \rangle )/2$ produced in the PCD process after a time of interaction $\tau_i= 2.881$, in terms of the temperature $T$ and for different values of the mismatch parameter $y=0$ (dashed-dot line), $y=0.5$ (dashed line) and $y=0.7$ (dotted line). It is also displayed the invariant $\sqrt{S_0} $ (solid line) in terms of the temperature. The first three lines intersect $\sqrt{S_0} $ at different values of the temperature $T$. Computing the temperature at the intersection points we find out that they coincide with the corresponding critical temperature $T_c$ for each $y$. For temperatures such that the average number of photons is in the shaded region of the figure, the system is entangled. Under the solid curve the average number of correlated photons produced in the interaction are not enough to attain the {\it break-even}, given by the square root of the invariant $S_0$, this is, they are not enough to exceed the uncorrelated photons present in the initial thermal state, and the system is separable. 

We can then define a simple entanglement witness 
\begin{equation}\label{witness}
{\cal W} (T)=  \sqrt{{ \bar n}_{10}(T){ \bar n}_{20}(T) ( { \bar n}_{10}(T)+1) ( { \bar n}_{20}(T)+1)}  - \langle {\hat n}(T)   \rangle 
\end{equation}
As usual, the negative value of ${\cal W} (T)$ indicates entanglement. To check wether the two modes are entangled, it will be enough to measure the average number $\langle {\hat n} (T)  \rangle $ of correlated photons generated in the two modes and the equilibrium temperature $T$ of the nonlinear medium, since this define the initial average number of photons ${ \bar n}_{j0}(T), (j=1,2)$ in each mode.

\section{Summary and concluding remarks}

We have computed the joint effect of a phase mismatch and of the temperature in the degree of entanglement in PDC processes described by Hamiltonean (\ref{hamiltonean}).  

Using the time invariants associated to that hamiltonean we computed the minimum interaction time needed before entanglement appears and have shown how it behaves with the phase mismatch and temperature. 

We have also shown that for each phase mismatch there is a critical temperature, above which the entanglement disapears. This study enables us to draw the border line between separable and entangled states in the plane of the parameters, phase mismatch and temperature. Depending on the interaction time, and on the phase mismatch it is possible to obtain entanglement at room temperature. Finally we propose  a feasible  entanglement witness.

 \section{Appendix}
 
 The Heisenberg equations of motion for modes $1$ and $2$ are
\begin{equation}\label{Heis}
\frac{d}{dt} 
\left( \begin{array}{cc}
\hat{a}_1\\
\hat{a}_2^{\dag}\\
\end{array}
\right)
= \left(
\begin{array}{cc}
-i \omega_1  & ig e^{- i [2 (\omega   + \delta) t  + \varphi ]}\\
-ig e^{ i [2 (\omega   + \delta) t + \varphi] }& i \omega_2 \\
\end{array} 
\right)
\left(
\begin{array}{c}
\hat{a}_1\\
\hat{a}_2^{\dag}\\
\end{array} 
\right)
\end{equation}
jointly with their hermitian conjugates, which are decoupled from these two. In order to obtain an autonomous system of equations we define new annhilation operators $\hat{b}_1$ and $\hat{b}_2$, by the canonical transformation
\begin{equation}\label{FA}
\left(\begin{array}{c}
\hat{b}_1\\
\hat{b}_2^{\dag}\\
\end{array}
\right)=
\left(\begin{array}{cc}
e^{ i (\omega   + \delta) t}& 0\\
0&e^{- i (\omega   + \delta) t}\\
\end{array}
\right)
\left(\begin{array}{c}
\hat{a}_1\\
\hat{a}_2^{\dag}\\
\end{array}
\right)
\end{equation}
Physically this corresponds to move to a reference frame that rotates with the frequency $(\omega   + \delta)$.
The Heisenberg equations for the new  operators $\hat{b}_1$ and $\hat{b}^{\dag }_2$ are then given by
\begin{equation}\label{Heis1}
\frac{d}{dt}
\left(
\begin{array}{c}
\hat{b}_1\\
\hat{b}_2^{\dag}\\
\end{array}
 \right)
= \left(
\begin{array}{cc}
i (\omega - \omega_1+ \delta)& ig e^{- i  \varphi} \\
-ige^{i  \varphi} & i (\omega - \omega_1- \delta)\\
\end{array}
\right)
\left(\begin{array}{cc}
\hat{b}_1\\
\hat{b}_2^{\dag}\\
\end{array} 
\right)
\end{equation}
They are linear time independent differential equations and can be readily integrated. The only solution of eq.(\ref{Heis1}) with physical interest is given by eqs.(\ref{time1}, \ref{time2}) of Section 2, and corresponds to the condition $ \delta < g $ where the bandwidth is small compared with the coupling parameter. 

It is immediate to prove that 
\begin{equation}\label{Inv}
\langle {\hat n}_1  (t)  \rangle  - \langle {\hat n}_2 (t)  \rangle =  { \bar n}_{10}  -  { \bar n}_{20}
\end{equation}
is an integral of motion. This is a generalization, for nonzero mismatch, of the analog integral of motion derived in \cite{LOUISELL1961}.

Using the time dependent bosonic operators given in eqs.(\ref{time1}) and (\ref{time2}) we compute the entries of the covariance matrix $\sigma$ for the thermal initial state
\begin{equation}\label{sigmajj}
\sigma_{jj} (\tau)   = \langle {\hat n}_j  (\tau)  \rangle  +\frac{1}{2}  \,\,\ ; \,\,\,\  (j=1,2)
\end{equation}
\begin{equation}\label{sigma12}
\sigma_{12} (\tau)   = \sigma_{21} (\tau)  = \sigma_{34} (\tau)  = \sigma_{43} (\tau)  =0  \,\,\ ; \,\,\,\  (j=1,2)
\end{equation}
\begin{equation}\label{sigma13}
\sigma_{24} (\tau)   = \sigma_{13} (\tau)   \,\,\,\ ; \,\,\,\    \sigma_{23} (\tau)   = \sigma_{14} (\tau)
\end{equation}
\begin{equation}\label{sigma13}
\sigma_{13} (\tau)  = ({ \bar n}_{10} + { \bar n}_{20} +1) \frac{S(x \tau)}{x} \left[ \sin ( \theta \tau) C(x \tau) \\
-  \frac{\cos ( \theta \tau )y}{x}S (x \tau)  \right]
\end{equation}
\begin{equation}\label{sigma14}
\sigma_{14} (\tau)   = ({ \bar n}_{10} + { \bar n}_{20} +1) \frac{S(x \tau)}{x} \left[ \cos (  \theta \tau) C(x \tau)\\
+  \frac{\sin ( \theta  \tau )y}{x} S (x \tau)  \right] 
\end{equation}
where $\theta =(\omega_1^{\prime}   + \omega_2^{\prime} )$ and $\omega_j^{\prime} = {\bar \omega}_j  +y$, $j=1,2$.

The CM for the vacuum state is easily obtained, by making ${ \bar n}_{10}={ \bar n}_{20}=0$ in the above equations.  
\\\\\\

\end{document}